\documentclass[aps,pra,reprint,groupedaddress,amsmath,amssymb]{revtex4-1}
\usepackage{bm}
\usepackage{graphicx}
\usepackage[usenames, dvipsnames]{color}
\usepackage{subcaption}
\graphicspath{{figures/}}

\newcommand*\diff{\mathop{}\!\mathrm{d}}

\begin{document}
\title{Hybrid master equation for calorimetric measurements}%

\author{Brecht Donvil}%
\email{brecht.donvil@helsinki.fi}
\affiliation{Department of Mathematics and Statistics, University of Helsinki, P.O. Box 68, 00014 Helsinki, Finland}
\author{Paolo Muratore-Ginanneschi}%
\email{paolo.muratore-ginanneschi@helsinki.fi}
\affiliation{Department of Mathematics and Statistics, University of Helsinki, P.O. Box 68, 00014 Helsinki, Finland}
\author{Jukka P. Pekola}%
\email{jukka.pekola@aalto.fi}
\affiliation{Low Temperature Lab (OVLL), Aalto University School of Science, P.O. Box 13500, 00076 Aalto, Finland}%
\date{November, 2018}
\begin{abstract}
Ongoing experimental activity aims at calorimetric measurements of thermodynamic indicators of quantum integrated systems. We study a model of a driven qubit in contact with a finite-size thermal electron reservoir. The temperature of the reservoir changes due to energy exchanges with the qubit and an infinite-size phonon bath. Under the assumption of weak coupling and weak driving, we model the evolution of the qubit-electron-temperature as a hybrid master equation for the density matrix of the qubit at different temperatures of the calorimeter. We compare the temperature evolution with an earlier treatment of the qubit-electron model, where the dynamics were modelled by a Floquet master equation under the assumption of drive intensity much larger than the qubit-electron coupling squared. We numerically and analytically inquire the predictions of the two mathematical models of dynamics in the weak-drive parametric region. We numerically determine the parametric regions where the two models of dynamics give distinct temperature predictions and those where their predictions match.
\end{abstract}
\maketitle

\section{Introduction}
One is often interested in deriving a reduced dynamics from the full microscopic description of a system, for example of a subsystem or of a macroscopic variable. In deriving the reduced dynamics, usually a set of approximations is made on parameters of the system. For a concrete physical situation, however, it is not always clear for which values of the parameters the derived dynamics are valid. 


 In this manuscript we compare two different reduced dynamics of a concrete system.  
Our motivation is an experiment proposed by \cite{PeSo13}, that aims to measure thermodynamic indicators of a driven qubit system in contact with a thermal environment. 
For a detailed discussion of the experiment we refer to \cite{PeSo13, DonvilCal}. In essence, the setup by \cite{PeSo13} is a nanoscale electric circuit, containing a superconducting qubit and a resistor element. The electrons in the resistor form a calorimeter, a finite sized environment of the qubit. The size of the resistor should be small enough, such that temperature fluctuations due to energy exchanges between the qubit and calorimeter are detectable.

The proposal by \cite{PeSo13} was previously modelled by \cite{KuMu16} as a stochastic jump process for the state of the qubit and the temperature of the calorimeter. The authors of \cite{DonvilCal} introduced electron-phonon coupling to the model. This interaction leads to additional drift and diffusion terms in the evolution of the temperature of the electrons. Secondly, the authors supposed a strong periodic drive. Under this condition they used the stochastic jump equation  derived by \cite{BPfloquet}, which is based on Floquet's theorem, to model the dynamics of the qubit. In the Floquet stochastic jump equation, the dynamics of the qubit are expressed in terms of solutions of the non-interacting periodically driven qubit. This modeling of the dynamics is convenient to study the evolution of the electron temperature numerically and analytically. The authors of \cite{DonvilCal} expressed the qubit-calorimeter dynamics in terms of Chapman-Kolmogorov type master equation. In what follows, we refer to this equation as the Floquet master equation. One of the main results of \cite{DonvilCal} was to derive from the Floquet master equation a Fokker-Planck equation for the probability distribution of the electron temperature on long time scales, by eliminating the underlying qubit dynamics. 

The disadvantage of using the Floquet stochastic equation is that it requires, additionally to the usual set of assumptions required for the Born-Markov approximation \cite{open, Rivas}, that the strength of the drive is much larger than the qubit-calorimeter coupling  squared \cite{BPfloquet}. This might not always be the physical reality in experiments.
In the current paper, we aim to study the temperature behaviour with a different model for the dynamics of the qubit. We use an unravelling \cite{Dalibard} of the usual Lindblad equation where the drive is added as a perturbation to the non-dissipative part. Besides the assumptions made for the Born-Markov approximation, this approach requires the strength of the drive to be much smaller than the level spacing of the qubit.

One of our main results is the description of the qubit-calorimeter as a hybrid master equation of the form \cite{ChKoMaSu2011,Di2014}. A hybrid master equation describes the joint evolution of quantum and classical variables, in the present case of the qubit wave function and the temperature of the calorimeter. The quantum discord related to temperature measurements of qubit-temperature states evolving according to the hybrid master equation is zero. This tells us that within our model by measuring the temperature, we can not detect any quantumness \cite{OllZu2002}. 

If the qubit is driven long enough, the qubit-calorimeter reaches a steady state: the electron temperature fluctuates around a stationary temperature $T_S$. On this time-scale, it is possible to derive from the hybrid master equation an effective equation for the temperature evolution. The effective equation has the form of a time-autonomous Fokker-Planck equation. The the qubit dynamics can be eliminated with the use of multi time-scale perturbation theory. We numerically compare predictions of the hybrid and Floquet master equations. We identify the region of parameters in which the Floquet and weak drive dynamics give the same temperature predictions. We find for which values of the qubit-calorimeter coupling and drive strength both predict the same value for $T_S$.  Experimentally this a good indicator: measuring the average temperature is far easier than measuring the fluctuations.

The structure of the paper is as follows.
In Section \ref{sec:mod} we shortly introduce the qubit-calorimeter model. We recap results by \cite{KuMu16} to describe the evolution of the qubit-calorimeter as a qubit state-temperature process. In Section \ref{sec:Me} we describe the qubit-temperature process as a hybrid master equation for the qubit-temperature density operator. Section \ref{sec:Ef} is devoted to deriving a Fokker-Planck equation for the temperature under the assumption of resonant driving of the qubit. On the hybrid master equation we perform time-multiscale perturbation theory in order to average out the qubit dynamics.
In Section \ref{sec:sim} we numerically study the qubit-calorimeter model numerically and compare the results to those obtained from the Floquet modelling of the qubit dynamics. Finally, in section \ref{sec:con} we shortly discuss the results.

\section{Qubit-Calorimeter model}\label{sec:mod}
We provide a short description of the qubit calorimeter model as proposed by \cite{PeSo13}. The setup consists of a driven qubit in contact with a finite-size electron bath on varying temperature $T_e$. The electron bath itself is in contact with an infinite-size thermal bath of phonons, on temperature $T_p$. The full Hamiltonian is 
\begin{equation}
H=H_q(t)+H_{qe}+H_e+H_{ep}+H_p
\end{equation}
The Hamiltonian of the qubit is 
\begin{equation}
H_q(t)=\frac{\hbar \omega}{2} \sigma_z+\kappa\hbar \omega(e^{-i\omega_d t}\sigma_++e^{i\omega_d t}\sigma_-),
\end{equation}
where $\sigma_z$ denotes the canonical Pauli matrix and $\omega_d$ is the driving frequency.
The interaction between the qubit and electrons is described by
\begin{equation}
H_{qe}=g\frac{\sqrt{8\pi}\epsilon_F}{3N}\sum_{k\neq l \in \mathbb{S}} (\sigma_++\sigma_-) a^\dagger_ka_l.
\end{equation}
$\sigma_+$ and $\sigma_-$ are the creation and annihilation operator for the qubit and $a_k$, $a^\dagger_k$ for the electrons. The sum is restricted to an energy shell $\mathbb{S}$ close to the Fermi energy $\epsilon_F$ of the electrons. 
$H_e$ and $H_p$ are the free electron and phonon Hamiltonians and $H_{ep}$ is the Fr\"{o}hlich interaction term between them \cite{Fr1952}. We will not explictly study the electron-phonon interaction in this work. Earlier works \cite{lif,well,PeBa2017} have shown that it induces a drift on the electron temperature towards the phonon temperature and noise.

In order to formulate an evolution equation for the qubit-calorimeter system, it is important to discuss the time scales involved in the model. The fastest time scale in the model is the relaxation time of the electrons to a thermal state \cite{PoGu1997} $\tau_{ee}\sim 1$ ns. The electron-phonon interaction takes place on a time scale $\tau_{ep}\sim 10^4$ ns \cite{GaVi2015} and the relaxation time of the qubit is typically up to $\tau_R\sim 10^5$ ns \cite{WaAx2015}. The large timescale separation $\tau_{ee}\ll \tau_R$ allows us to invoke the Markov approximation. Additionally we assume that the characteristic time scale $\tau_{eq}$ of the qubit-calorimeter interaction satisfies $\tau_{eq}\ll \tau_{ep}\ll\tau_R$. Under this assumption we can evaluate the qubit transitions rates using the Fermi-Dirac distribution for the electron bath.

Under the above approximations we express the qubit dynamics in terms of a stochastic Schr\"{o}dinger equation, which consists of a continuous evolution interrupted by sudden jumps
\begin{align}\label{eq:quj}
&\diff \psi(t) =\psi(t+\diff t)-\psi(t)\nonumber\\&\quad= -i H_q(t) \psi(t)\diff t -\frac{1}{2}\bigg(\Gamma_\downarrow (\sigma_+\sigma_--\|\sigma_-\psi(t)\|^2) \psi(t)\nonumber\\&\quad+\Gamma_\uparrow (\sigma_-\sigma_+-\|\sigma_+\psi(t)\|^2) \psi(t)\bigg)\diff t\nonumber\\
&\quad+\bigg(\frac{\sigma_-\psi(t)}{\|\sigma_-\psi(t)\|}-\psi(t)\bigg)\diff N_\downarrow+\bigg(\frac{\sigma_+\psi(t)}{\|\sigma_+\psi(t)\|}-\psi\bigg)\diff N_\uparrow.
\end{align}
Where $N_\downarrow$ and $N_\uparrow$ are Poisson counting processes.
We estimate the temperature dependence of the calorimeter temperature on its energy with the Sommerfeld expansion, see e.g. \cite{solid}. Under our assumptions, the energy $E$ of the calorimeter only changes on time scales much larger than $\tau_{eq}$. We find that 
\begin{equation}\label{eq:t1}
\diff T_e^2(t)= \frac{1}{N\gamma}\diff E(t),
\end{equation}
where 
\begin{equation}
\gamma=\frac{\pi^2 k_B^2}{4 \epsilon_F}.
\end{equation}
In our model we have two contributions to the change in energy of the calorimeter
\begin{equation}\label{eq:t2}
\diff E(t)=\diff E_{eq}(t) +\diff E_{ep}(t).
\end{equation}
The energy exchange due to interaction with the qubit is given by
\begin{equation}\label{eq:t3}
\diff E_{eq}(t)= \hbar\omega(\diff N_\downarrow -\diff N_\uparrow).
\end{equation}
The energy exchanged due to the electron phonon interaction can be modelled by a drift and diffusion term \cite{lif,well,PeBa2017}
\begin{equation}\label{eq:t4}
\diff E_{ep}(t)=\Sigma V (T^5_p-T^5_e)\diff t +\sqrt{10 k_B\Sigma V}T_p^3\diff w(t),
\end{equation}
where $V$ is the volume of the calorimeter, $\Sigma$ is a material constant and $T_p$ is the phonon temperature and $\diff w(t)$ is the increment of a Wiener process.
The Poisson processes $N_\downarrow$ and $N_\uparrow$ are characterised by the conditional expectation values
\begin{equation}\label{eq:diffup}
\mathsf{E}(\diff N_{\uparrow}|\psi,T)=\Gamma_{\uparrow}\|\sigma_{+}\psi\|^2\diff t.
\end{equation}
\begin{align}\label{eq:diffdown}
\mathsf{E}(\diff N_{\downarrow}|\psi,T)=&\Gamma_{\downarrow}\|\sigma_{-}\psi\|^2\diff t
\end{align}
The decay rate is defines as
\begin{subequations}\label{eq:rates}
\begin{equation}
\Gamma_{\downarrow}=\left\{
                \begin{array}{ll}
                  \dfrac{g^2 \omega\, e^{ \hbar \omega/k_B T_e}}{e^{\hbar \omega/k_B T_e}-1} \quad &\textrm{for} \quad T^2_e>\frac{\hbar\omega}{N\gamma}\\
                  1&\textrm{for}\quad \frac{\hbar\omega}{N\gamma}\geq  T^2_e>0
                \end{array}\right.
\end{equation} 
and the excitation rate equals
\begin{equation}
\Gamma_{\uparrow}=\left\{
                \begin{array}{ll}
                  \dfrac{g^2 \omega }{e^{\hbar \omega/k_B T_e}-1} \quad&\textrm{for} \quad T^2_e>\frac{\hbar\omega}{N\gamma}\\
                 0&\textrm{for}\quad\frac{\hbar\omega}{N\gamma}\geq T^2_e>0
                \end{array}\right.
\end{equation} 
\end{subequations}
The excitation rate is set to zero for temperatures squared lower than $\hbar\omega/N\gamma$. For such temperatures an excitation of the qubit would give a negative temperature: the calorimeter does not have enough energy. In our numerical studies of the model, we never actually reach these low temperatures.

From an experimental point of view one is mainly interested in the evolution of the temperature. In the next sections we show that on longer time scales of many periods of driving it is possible to derive an effective evolution equation for the temperature. 

\section{Hybrid master equation} \label{sec:Me}
In order to eliminate the qubit process from the qubit-temperature evolution, it is convenient to first express the dynamics of the qubit-temperaturd in terms of a master equation. We define the process for the temperature squared $\xi(t)=T_e^2(t)$: Combining equations \eqref{eq:t1}-\eqref{eq:t4}, $\xi(t)$ obeys the stochastic differential equation 
\begin{align}\label{eq:tempj}
\diff \xi(t)=&\frac{1}{N\gamma}\bigg(\hbar\omega(\diff N_\downarrow -\diff N_\uparrow)+\Sigma V (T^5_p-\xi^{5/2}(t))\diff t \nonumber\\&+\sqrt{10 k_B\Sigma V}T_p^3\diff w_t\bigg).
\end{align} Let 
\begin{align}
&P(\psi,\psi^*,X,t)\nonumber\\&=P(X\leq \xi(t) < X+\diff X \textrm{ and qubit in state } \psi)
\end{align} be the probability for a qubit to be in a state $\psi$ and the calorimeter to have temperature squared $X$ at time $t$. In Appendix \ref{ap:bound}, we derive a master equation for $P$ and discuss the relative boundary conditions. For our purpose, however, it is more convenient to work with a different object than the full probability distribution. Let us first define the marginal temperature-squared distribution is defined as 
\begin{equation}
F(X,t):=\int \textrm{D} \psi\textrm{D}\psi^*\,\langle \psi| \mathbb{I}|\psi\rangle\, P(\psi,\psi^*,X,t).
\end{equation}
Additionally, we introduce a notation for the expectation values of the canonical Pauli matrices at temperature squared $X$
\begin{equation}
\langle \sigma_i\rangle_X :=\int \textrm{D}\psi\textrm{D}\psi^*\langle \psi| \sigma_i|\psi\rangle P(\psi,\psi^*,X,t)
\end{equation}
with $i=z,\, y,\, z$.
Using the above definitions we define the qubit density operator at temperature squared $X$ as
\begin{equation}\label{eq:rho}
\rho(X,t):=\frac{1}{2}(F(X,t) \,\mathbb{I}+\langle\vec{\sigma} \rangle_X\cdot\vec{\sigma}).
\end{equation}
In Appendix \ref{ap:lin} we show that $\rho(X,t)$ satisfies the master equation 
\begin{align}\label{eq:fullrho}
\frac{\diff \rho}{\diff t}(X,t)=\frac{1}{N}\mathcal{L}_X \rho(X,t)+\rho(X,t)+M(\rho)(X,t)
\end{align}
with 
\begin{equation}
\mathcal{L}_X\rho(X,t)=\mathcal{L}^{(1)}_X \rho(X,t)+\mathcal{L}^{(2)}_X \rho(X,t)
\end{equation}
\begin{equation}
\mathcal{L}^{(1)}_X \rho(X,t)=-\frac{\Sigma V}{N\,\gamma}\partial_X\big((T_p^{5}-X^{5/2})
\rho(X,t)\big),
\end{equation}
\begin{equation}
\mathcal{L}^{(2)}_X \rho(X,t)=\frac{(\sqrt{10\Sigma V k_B}T_p^3 )^2}{2\,N^{2}\,\gamma^{2}}\partial_X^2\rho(X,t)
\end{equation}
and
\begin{align}
&M(\rho)(X)=-\frac{i}{\hbar} [H(t),\rho(X)]\nonumber\\&+ G_\downarrow\left(X-\frac{\hbar\omega}{N\gamma}\right)\sigma_-\rho\left(X-\frac{\hbar\omega}{N\gamma}\right)\sigma_+ \nonumber \\&+ G_\uparrow\left(X+\frac{\hbar\omega}{N\gamma}\right)\sigma_+\rho\left(X+\frac{\hbar\omega}{N\gamma}\right)\sigma_-\nonumber\\&-\frac{1}{2}
G_\downarrow(X)\{\sigma_+\sigma_-,\rho(X)\}-\frac{1}{2}G_\uparrow(X)\{\sigma_-\sigma_+,\rho(X)\} ,
\end{align}
where we have defined
\begin{subequations}\label{eq:rates2}
\begin{equation}
G_\downarrow(X)=\frac{g^2 \omega\, e^{ \hbar \omega/k_B \sqrt{X}}}{e^{\hbar \omega/k_B \sqrt{X}}-1}\theta\bigg(X-\frac{\hbar\omega}{N\gamma}\bigg)+\theta\bigg(\frac{\hbar\omega}{N\gamma}-X\bigg).
\end{equation}
\begin{equation}
G_\uparrow(X)=\frac{g^2 \omega }{e^{\hbar \omega/k_B \sqrt{X}}-1}\theta\bigg(X-\frac{\hbar\omega}{N\gamma}\bigg).
\end{equation}
\end{subequations}
in accordance with the jump rates \eqref{eq:rates}, to explicitly show the dependency on $X$.

Equation \eqref{eq:fullrho} is a hybrid master equation, it describes the joint evolution of the classical variable $X$ and the quantum variable $\psi$. When the size of the calorimeter goes to infinity, $N\uparrow\infty$, qubit variables at different temperatures get decoupled and the above equation reduces to an ordinary Lindblad equation for a qubit interacting with a thermal environment.

 In Appendix \ref{ap:hybrid} we show that our equation can be identified with a hybrid master equation of the form discussed in \cite{ChKoMaSu2011,Di2014}. From one of the results in \cite{ChKoMaSu2011}, we deduce that when only the temperature is measured, the quantum discord of a state $\rho(X)$ is zero. Quantum discord is defined as the difference between two classically equivalent expressions for mutual information \cite{OllZu2002}. It is an indicator for the quantumness of the correlations obtained from measuring the temperature. Equation \eqref{eq:fullrho} is not of the form of a classical Pauli master equation, as is the case for the Floquet approach \cite{DonvilCal}. Nevertheless, from the quantum discord being zero, we can conclude that by measuring the temperature, we cannot detect quantum effects.
\section{Effective temperature process}\label{sec:Ef}
Let us now assume the existence of a separation of time scales in the model, namely, that temperature of the calorimeter equilibrates much slower than the qubit does. Concretely, we will expand the dynamics under the limit of an infinite size calorimeter, and work a time scale on which the temperature evolves and the qubit has already relaxed. The expansion parameter is the inverse of the amount of electrons $N$ in the calorimeter
\begin{equation}
\varepsilon=\frac{1}{N}.
\end{equation}
The qubit dynamics take place on the time scale set by $t$, for the temperature dynamics we introduce the second time 
\begin{equation}
\tau=\epsilon t.
\end{equation}
When we write the dependence of the density operator $\rho$ on the two scales
\begin{equation}
\rho(X,t)=\tilde{\rho}(X,t,\tau),
\end{equation}
the time derivative becomes
\begin{equation}\label{eq:fulldt}
\frac{\diff \rho}{\diff t} (X,t)=\partial_t \tilde{\rho}(X,t,\tau)+ \epsilon\partial_\tau \tilde{\rho}(X,t,\tau).
\end{equation}
To perform the perturbative expansion, we assume that the process has already relaxed on the shortest time scale. We consider the density operator
\begin{equation}\label{eq:rho2}
\bar{\rho}(X,\tau)=\lim_{t\uparrow+\infty}\tilde{\rho}(X,t,\tau).
\end{equation}
It convenient to write the matrix elements of $\rho(X)$
\begin{equation}\label{eq:defP}
\bar{\rho}(X,\tau)=\begin{pmatrix}
P_1(X,\tau)& P_3(X,\tau)\\
P_4(X,\tau)& P_2(X,\tau)
\end{pmatrix}.
\end{equation}
into a vector $\vec{P}(X,\tau)$. By the definition of $\rho(X,t)$ \eqref{eq:rho}, we can see that the off-diagonal elements
\begin{subequations}
\begin{equation}
P_3(X,\tau)=\frac{1}{2}\langle\sigma_x-i\sigma_y\rangle_X=\langle\sigma_-\rangle_X
\end{equation}
\begin{equation}
P_4(X,\tau)=\frac{1}{2}\langle\sigma_x+i\sigma_y\rangle_X=\langle\sigma_+\rangle_X
\end{equation}
are each others adjoint.
\end{subequations}
Expanding equation \eqref{eq:fullrho} in terms of $\varepsilon$ and using equation \eqref{eq:fulldt}, we get into
\begin{align}\label{eq:differentorders}
&\varepsilon\frac{\diff \vec{P}(X)}{\diff t}=\varepsilon\mathcal{L}^{(1)}\vec{P}(X) +\varepsilon^2\mathcal{L}^{(2)}\vec{P}(X) \nonumber\\&+ M^{(0)}(\vec{P})(X)+\sum_{n=1}\frac{\varepsilon^n}{n!}\partial_X^n\left( M^{(n)}(\vec{P})(X)\right).
\end{align}
with
\begin{equation}
M^{(0)}=\begin{pmatrix}
-G_\downarrow(X) & G_\uparrow(X)& i\lambda &-i\lambda\\
G_\downarrow(X) & -G_\uparrow(X)& -i\lambda &i\lambda\\
 i\lambda  & - i\lambda &-G(X)/2 &0\\
 - i\lambda  & i\lambda &0&-G(X)/2 
\end{pmatrix}.
\end{equation}
The sum of the rates is defined as 
\begin{equation}
G(X)=G_\downarrow(X)+G_\uparrow(X)
\end{equation} and the higher orders in the expansion of $M$ are
\begin{equation}
M^{(n)}=\left(\frac{\hbar\omega}{\gamma}\right)^n\begin{pmatrix}
0 & G_\uparrow(X) & 0 & 0\\
 (-1)^n G_\downarrow(X)  &0& 0 & 0\\
0 & 0 & 0 & 0\\
0 & 0 & 0 & 0\\
\end{pmatrix}.
\end{equation}
Note that the matrix $M^{0}$ corresponds to the Lindblad equation in the infinite calorimeter limit.

In Appendix \ref{app:Ef} we solve equation \eqref{eq:differentorders} at different orders in $\varepsilon$ using a Hilbert expansion \cite{Pavbook} of the probability distribution
\begin{equation}\label{eq:hilbertexp}
\vec{P}(X,t)= \sum_{n=0}^{+\infty} \epsilon^n \vec{P}^{n}(X,t).
\end{equation}
The marginal temperature distribution is obtained by taking the trace of $\bar{\rho}$, see equation \eqref{eq:rho}, which corresponds to summing the first two components of $\vec{P}$ \eqref{eq:defP}:
\begin{align}
F(X,\tau)=& P_1(X,\tau) +P_2(X,\tau) \nonumber\\=&\sum_{n=0}^\infty\varepsilon^n(P^{(n)}_1(X,\tau) +P^{(n)}_2(X,\tau))\nonumber\\=&\sum_{n=0}^\infty\varepsilon^n F^{(n)}(X,\tau).
\end{align}
The result of Appendix \ref{app:Ef} is an effective equation for $F(X,\tau)=F^{(0)}(X,\tau)+\varepsilon F^{(1)}(X,\tau)$ up to second order in in $\varepsilon$ for resonant driving
\begin{align}\label{eq:pertF}
&\partial_\tau F(X,\tau)=\nonumber\\&-\partial_X \bigg[\bigg(\frac{\Sigma V}{N\gamma}(T_p^5-X^{5/2}) + \frac{1}{N} j^{(1)}(X)\nonumber\\&+\frac{1}{N^2}j^{(2)}(X)\bigg) F(X,\tau) \bigg]+\partial_X^2\bigg[\bigg(\frac{(\sqrt{10\Sigma V k_B}T_p^3 )^2}{2\,N^{2}\,\gamma^{2}}\nonumber\\&+\frac{1}{N^2}\Delta^{(1)}(X)+\frac{1}{N^2}\Delta^{(2)}(X)\bigg) F(X,\tau) \bigg],
\end{align}
where we have defined the corrections to the drift as 
\begin{align}
j^{(1)}(X)=-\langle v_1,M^{(1)}Q\rangle
\end{align}
\begin{align}
&j^{(2)}(X)\nonumber\\&=-
\langle v_1|\partial_X\left(\frac{1}{\lambda_3\langle v_3 |w_3\rangle}|M^{(1)} w_3\rangle\langle v_3|\right) M^{(1)} Q\rangle\nonumber\\& -\langle v_1\partial_X\left(\frac{1}{\lambda_4\langle v_4|w_4\rangle}|M^{(1)}w_4\rangle\langle v_4|\right),M^{(1)} Q\rangle \nonumber\\&-\frac{1}{\lambda_4\langle v_4|w_4\rangle}\langle v_1|M^{(1)}w_4\rangle\langle v_4|(\mathcal{L}^{(1)})^\dagger Q\rangle\nonumber\\&-\frac{1}{\lambda_3\langle v_3|w_3\rangle}\langle v_1|M^{(1)}w_3\rangle\langle v_3|(\mathcal{L}^{(1)})^\dagger Q\rangle.
\end{align}
And the corrections to the diffusion coefficient
\begin{align}
\Delta^{(1)} (X)=\frac{1}{2}\langle v_1|M^{(2)}Q\rangle
\end{align}
\begin{align}
\Delta^{(2)}(X) =&-\frac{1}{\lambda_4\langle v_4|w_4\rangle}\langle v_1| M^{(1)}w_4\rangle\langle v_4|M^{(1)} Q\rangle\nonumber\\&-\frac{1}{\lambda_3\langle v_3|w_3\rangle}\langle v_1| M^{(1)}w_3\rangle\langle v_3|M^{(1)} Q\rangle.
\end{align}
The effective equation for the evolution of the temperature distribution \eqref{eq:pertF} has the form of a time-autonomous Fokker-Planck equation.

The stationary temperature $T_S$ is defined as the square root of $X_S$, for which the drift coefficient is zero.
The lowest order correction to the drift $j^{(1)}(X)$ explicitly allows us to estimate the dependence of the stationary temperature $T_S$ on the qubit-calorimeter coupling $g$ and the driving strength $\kappa$
\begin{equation}
j^{(1)}(X)=\frac{\hbar\omega^2g^2 4\kappa^2}{g^4\coth^2\left(\frac{\hbar\omega}{2k_B \sqrt{X}}\right)+8\kappa^2}.
\end{equation}
For large $\hbar\omega/(2k_B\sqrt{X})$, i.e. for small $X$, $\coth[ \hbar\omega/(2k_B\sqrt{X})]\approx 1$.
Under this approximation we find that
\begin{equation}
T_S\approx\bigg( T_p^5+\frac{1}{\Sigma V}\frac{\hbar\omega^2g^2 4\kappa^2}{g^4+8\kappa^2}\bigg)^{1/5}.
\end{equation}
Using the Floquet approach the $g$ dependence of $T_S$ was found to be \cite{DonvilCal}
\begin{equation}
T_S\approx\bigg( T_p^5+g^2\frac{O\big(\hbar\omega^2\big)}{\Sigma V}\bigg)^{1/5},
\end{equation}
where the weak dependence on the strength of the drive $\kappa$ is hidden in $O\big(\hbar\omega^2\big)$. For $g^2\ll \kappa$, the range in which the Floquet stochastic process is valid, both expressions show the same $g$-dependence.
\section{Simulations}\label{sec:sim}
We aim to compare temperature predictions by the weak-drive modelling of the qubit dynamics to those of the Floquet modelling studied in \cite{DonvilCal}. For the numerical integration of the dynamics, we take the similar parameters as \cite{DonvilCal}. The level spacing of the qubit is $\hbar\omega=0.5 k_B\times 1$ K, the volume of the calorimeter is $V=10^{-21} \textrm{m}^3$, $\Sigma= 2\times 10^{-9}$ W\, K\textsuperscript{-5}m\textsuperscript{-3} and $\gamma=1500k_B/$(1K). The strength of the drive $\kappa$ and the qubit-calorimeter coupling $g$ will be varied during the numerics.

Following the physical situation described in \cite{PeSo13}, at the start of the simulations the calorimeter and qubit are in thermal equilibrium with the phonon bath at temperature $T_p$. From the thermal distribution an initial state for the qubit is drawn. 

In order to numerically integrate the dynamics of the qubit-calorimeter system, time is discretized into steps of  the size $\diff t= (1000\omega_q)^{-1}$. The qubit state $\psi(t)$ and the electron temperature $T_e(t)$ is then updated from time $t$ to $t+\diff t$ in three steps: (1) the jump rates $\Gamma_{\uparrow/\downarrow}$ are calculated from $\psi(t)$ and $T_e(t)$. (2) a random number generator decides whether the system makes a jump. (3) $\psi(t+\diff t)$ and $T_e(t+\diff t)$ are calculated using equations \eqref{eq:quj} and \eqref{eq:tempj}. 

\begin{figure}
\centering
\includegraphics[scale=0.4]{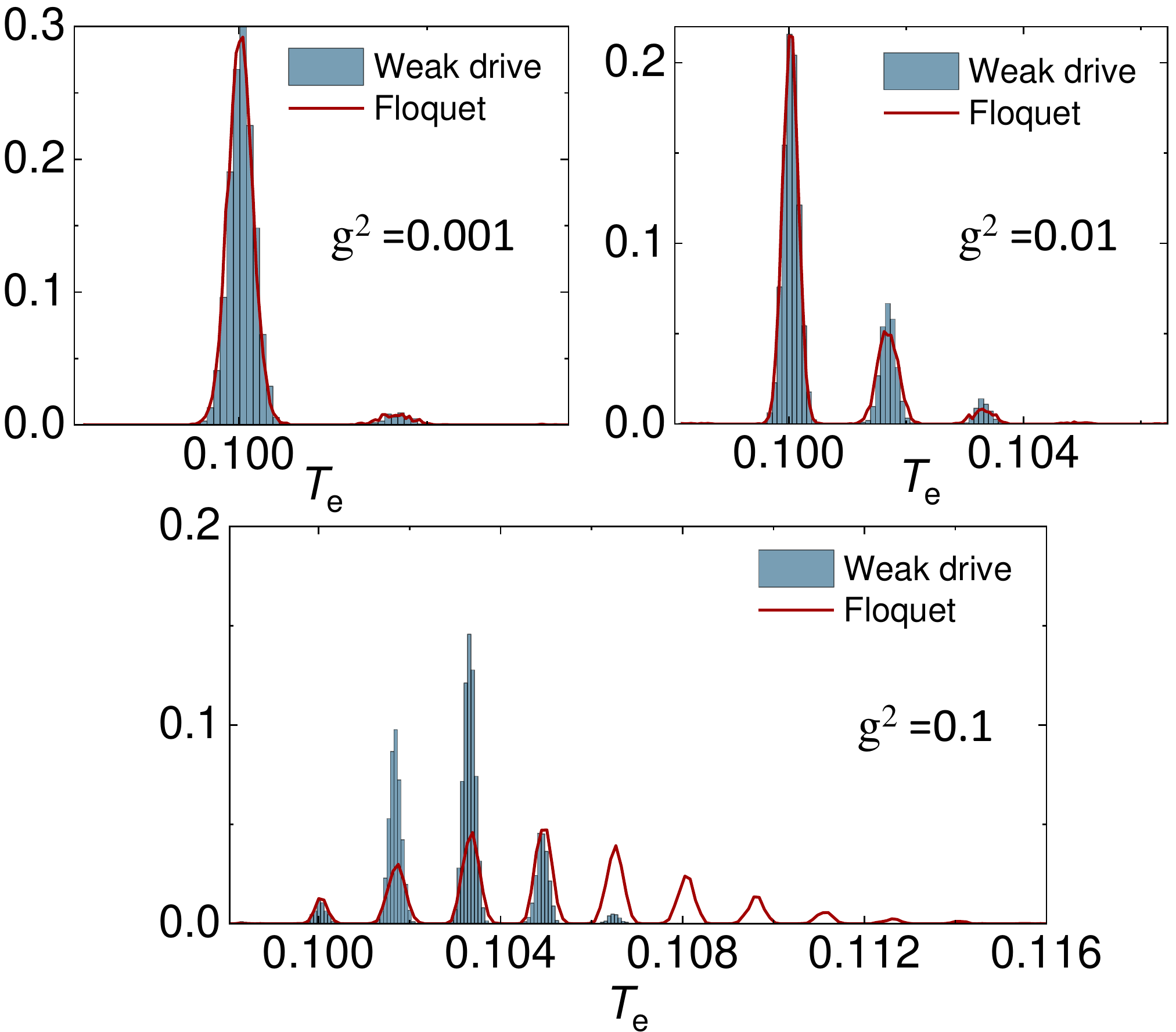}
\caption{Distribution of the electron temperature $T_e$ after a driving duration of 10$\times 2\pi/\omega$  for different values of the qubit-calorimeter $g$ and $\kappa=0.05$. The full (red) line is the distribution obtained from numerically integrating the qubit-temperature evolution with Floquet dynamics for the qubit,  the (blue) histogram comes from the weak drive dynamics \eqref{eq:quj} and \eqref{eq:tempj}, both from $10^4$ repetitions. The physical parameters used in the numerical integration are $\hbar\omega=0.5 k_B\times 1$ K, is $V=10^{-21} \textrm{m}^3$, $\Sigma= 2\times 10^{-9}$ W\, K\textsuperscript{-5}m\textsuperscript{-3} and $\gamma=1500k_B/$(1K).}
\label{fig:shortdistr}
\end{figure}

Figure \ref{fig:shortdistr} shows the temperature distribution of the calorimeter after a driving duration of 10 periods of the qubit $T=2\pi/\omega$. It is obtained from $10^4$ repetitions of numerically integrating equations \eqref{eq:quj} and \eqref{eq:tempj}. For low coupling strength $g^2$ the (red) line from the Floquet modelling overlaps with the (blue) histogram weak drive modelling.  When $g^2$ is increased the temperature distribution is shifted to the right, indicating that the assumptions required for the Floquet modelling of the qubit dynamics are broken. In this regime the latter overestimates the power exerted by the qubit. The explanation of the overestimate of the power resides in the assumption $\kappa\gg g^2$ needed to derive the Floquet jump equation.

\begin{figure}
\centering
\includegraphics[scale=0.7]{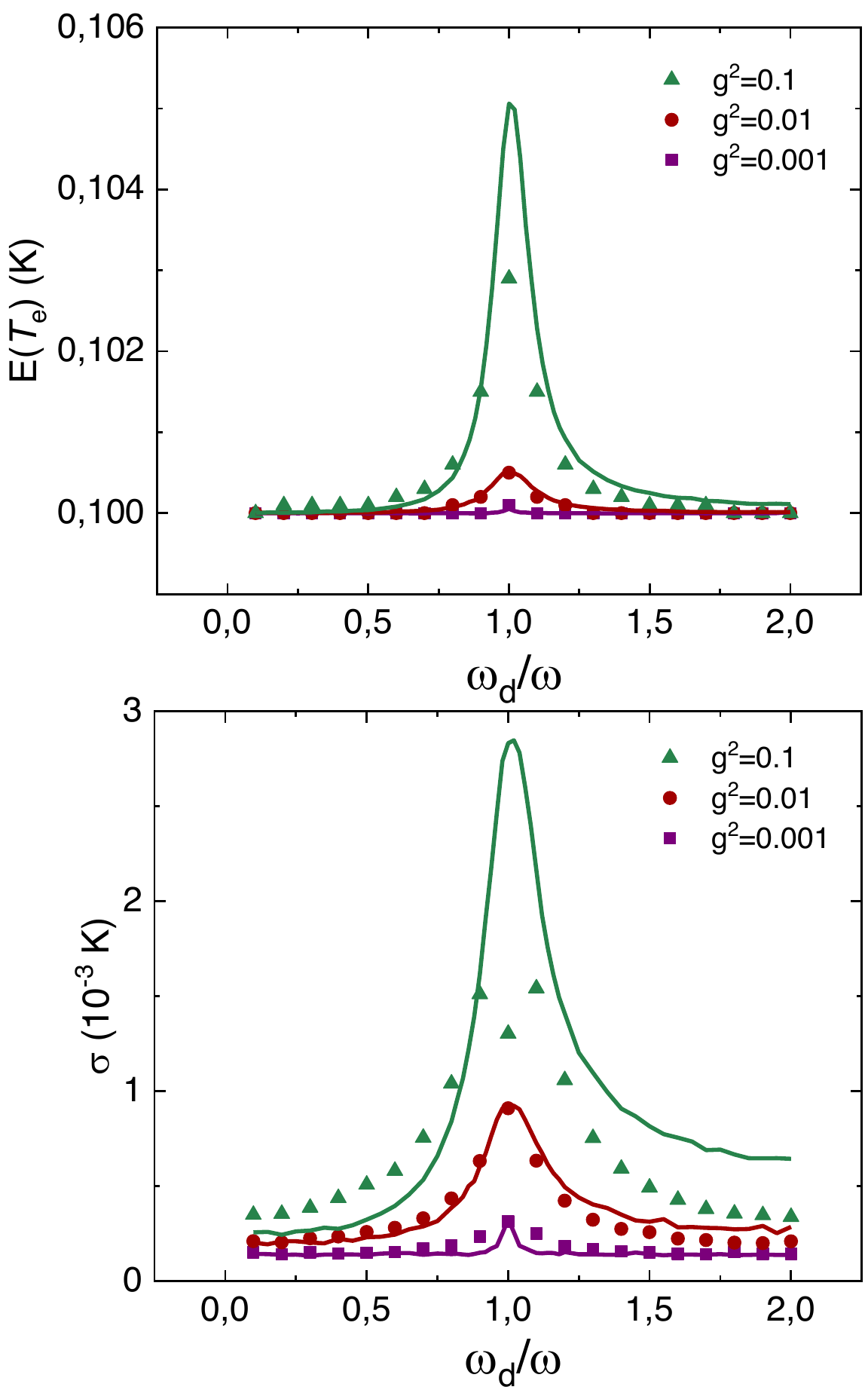}
\caption{ Mean (top) and standard deviation (bottom) of the temperature distribution after driving a duration of 10$\times 2\pi/\omega$, see Figure \ref{fig:shortdistr}, in function of the ratio of the driving and qubit frequency $\omega_d/\omega$ for different values of the qubit calorimeter coupling $g$. The stars are obtained from numerically integrating the qubit-temperature jump process with the weak-drive qubit-dynamics \eqref{eq:quj} and \eqref{eq:tempj}, and the full lines of numerically integrating with the Floquet modelling of the qubit-dynamics \cite{DonvilCal}. The parameters used for simulations are in the caption of Figure \ref{fig:shortdistr}.}
\label{fig:mshort}
\end{figure}

The mean and standard deviation of the temperature distribution in function of the ratio of the driving and qubit frequency $\omega_d/\omega$ after driving a duration of 10 periods of the qubit are shown in Figure \ref{fig:mshort}. The full lines are obtained from the Floquet modelling, while the points are from the numerics of the weak drive. Again we see that for low enough coupling $g^2$, the predictions from both modellings correspond well.

\begin{figure}
\centering
\includegraphics[scale=0.8]{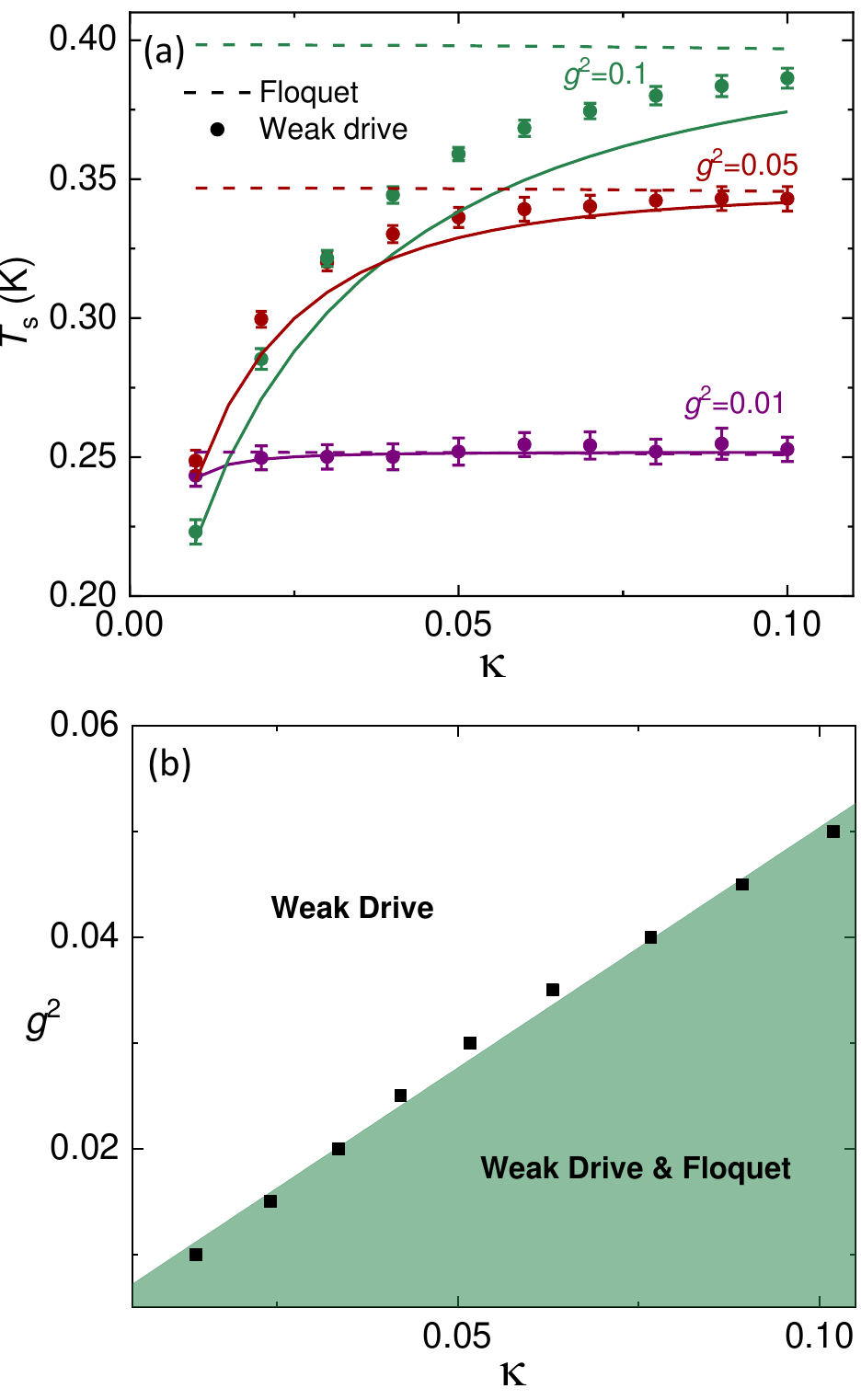}
\caption{A comparison of the stationary temperature $T_S$ of the calorimeter for the weak-drive and Floquet modeling of the qubit dynamics obtained  from numerics and analytics, \eqref{eq:pertF} and \cite{DonvilCal} respectively. (a) Dependence of the stationary temperature on the driving strength $\kappa$ for coupling $g$ held constant. The dots are obtained numerically from the weak drive and the full line from the analytics. The striped lines are analytic results of the Floquet modelling. (b) Parametric region for which the Floquet-modelling $T_S$ predictions correspond with the weak-drive-modelling predictions. In the green region both predict the same values, in the white region only they differ. The parameters used for simulations are in the caption of Figure \ref{fig:shortdistr}.}
\label{fig:ctk}
\end{figure}

Figure \ref{fig:ctk} (a) shows the mean temperature of the calorimeter, after it has reached a steady state. The mean is an estimate for the stationary temperature $T_S$ of the effective Fokker-Planck equation \eqref{eq:pertF}. The estimate of the stationary temperature is shown as a function of the driving strength $\kappa$ for different qubit-calorimeter coupling $g^2$ values. The value of $T_S$ predicted by the weak-drive modelling of the qubit dynamics asymptotically reaches the Floquet-modelling prediction. The parametric region we consider $\kappa\sim O(10^{-2})$ and $g^2\sim O(10^{-1})$ is in the the range of week driving. For large enough driving strength $\kappa$ compared to $g^2$, both approaches predict the same value for $T_S$. This indicates that the assumption $\kappa\gg g^2$ required for the Floquet modelling is met. By using $T_S$ as an indicator, we can estimate the parametric region of validity for the Floquet modelling of the qubit dynamics. 
Figure \ref{fig:ctk} (b) shows that estimated region. The region where the weak-drive dynamics estimate for $T_S$ plus one standard deviation obtained from numerics exceeds the predicted value by the Floquet modelling is coloured. This corresponds to the points in Figure \ref{fig:ctk} where the errorbars on the dots exceed the striped assymptotes. The slope between the two regions is $0.454\pm 0.013$, which means that $\kappa$ has to be about twice as large as $g^2$ for the Floquet modelling to hold. The experiment proposed by \cite{PeSo13} has the aim to measure the temperature of the bath. By measuring the steady state temperature at different levels of driving strength, it is possible to detect in which regime the Floquet approach holds for the experimental setup.

\begin{figure}
\centering
\includegraphics[scale=0.45]{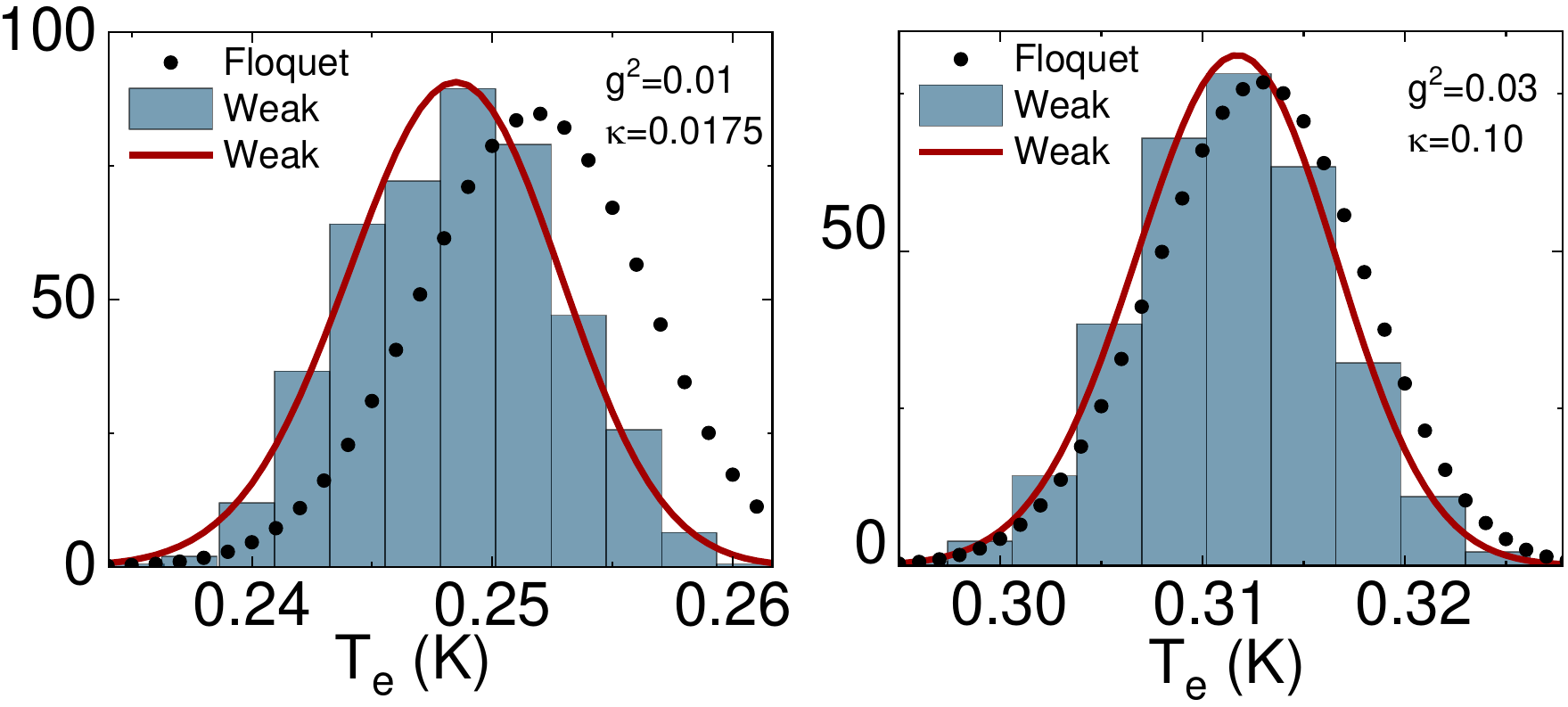}
\caption{Steady state distribution obtained from the numerics (blue histogram) compared to steady state predicted by the Fokker-Planck equation for the weak drive \eqref{eq:pertF} (red) line and the Floquet Fokker-Planck equation (black) circles \cite{DonvilCal}. The parameters for the left distribution are outside of the region of validity for the Floquet modelling, see Figure \ref{fig:ctk} (b). The parameters of the right figure are inside of this region.}
\label{fig:ssdistr}
\end{figure}

Figure \ref{fig:ssdistr} shows the temperature steady state distribution obtained from the numerics. It is compared to the steady state distribution of the Fokker-Planck equation \eqref{eq:pertF} and the Fokker-Planck equation from the Floquet modelling \cite{DonvilCal}. The distributions in the right figure correspond well, the values of $\kappa$ and $g^2$ are inside the green region in Figure \ref{fig:ctk} (b).
\section{Conclusion}\label{sec:con}
The evolution of the state and the qubit and temperature of the calorimeter can be formulated in terms of a hybrid master equation \eqref{eq:fullrho}. The quantum discord related to temperature measurements of qubit-temperature states as defined in \eqref{eq:rho} is zero. Which tells us that, although the hybrid master equation does not have a classical form, temperature measurements cannot detect any quantum effects.

Using time-multiscale perturbation theory, we were able to reduce the full qubit-temperature process to an effective Fokker-Planck equation for the electron temperature \eqref{eq:pertF}. 

We compared the numerical and analytic results for the weak-drive modelling of the qubit dynamics with the Floquet modelling studied in \cite{DonvilCal}. When the qubit calorimeter coupling squared $g^2$ is small enough compared to $\kappa$, temperature predictions from both correspond well quantitatively. For a few periods of driving they show the same temperature distributions. For long periods of driving, both modellings predict similar steady state statistics. When the qubit-calorimeter coupling is increased, such that the assumptions required for the Floquet approach are violated, the predictions of the models become quantitatively different. The Floquet modelling over-estimates the power exerted from the qubit to the calorimeter. This can be explained by the observation that in the derivation of the Floquet stochastic jump equation, it is assumed that the driving strength is much larger than the qubit-calorimeter coupling squared. Using the steady state temperature as an indicator, we estimated the regime of validity of the Floquet modelling of the qubit dynamics. In the experimental setup of \cite{PeSo13} this region can be directly measured.
\section{Acknowledgements}
The work of B. D. is supported by DOMAST.  B.~D. and P.M-G. also acknowledge support by the Centre  of Excellence  in  Analysis  and Dynamics  of  the Academy  of Finland. The work of J. P. P. is funded through Academy of Finland grant 312057 and from the European Union's Horizon 2020 research and innovation programme under the European Research Council (ERC) programme (grant agreement 742559).
\appendix
\section{Boundary conditions}\label{ap:bound}
For equations \eqref{eq:quj} and \eqref{eq:tempj} we impose reflective boundary conditions at $X=0$. 
We derive the master equation for the probability, let $f(\psi,\psi^*,X)$ be a smooth function, the time derivative of its average is 

\begin{align}\label{eq:bounds}
&\frac{\diff}{\diff t}\mathsf{E}(f)\nonumber\\&=\int_0^{+\infty}\diff X \int \mathrm{D}\psi \mathrm{D}\psi^*\bigg( \mathcal{L}^\dagger(f)(\psi,\psi^*,X)P(\psi,\psi^*,X)\nonumber\nonumber\\&\quad+K(\psi)\partial_\psi (f)(\psi,\psi^*,X)P(\psi,\psi^*,X)\nonumber\\&\quad+K^*(\psi)\partial_{\psi^*} (f)(\psi,\psi^*,X)P(\psi,\psi^*,X)\nonumber\\&\quad+\bigg[f\left(\phi_+,\bar{\phi}_+,X-\frac{\hbar\omega}{N\gamma}\right)-f(\psi,\psi^*,X)\bigg]\nonumber\\&\quad\quad\times G_\uparrow(X)\|\sigma_+\psi\|^2P(\psi,\psi^*,X,t)
\nonumber\\&\quad+\bigg[f\left(\phi_-,\bar{\phi}_-,X+\frac{\hbar\omega}{N\gamma}\right)-f(\psi,\psi^*,X)\bigg]\nonumber\\&\quad\quad\times G_\downarrow(X)\|\sigma_-\psi\|^2P(\psi,\psi^*,X,t)\bigg)
\end{align}
with $\phi_\pm$ the energy eigenstates of $\sigma_z$ with eigenvalues $\pm 1$, $G_{\uparrow/\downarrow}$ is defined in \eqref{eq:rates2} and 
\begin{align}\label{eq:ok}
K(\psi)=&-i H_q(t) \psi\diff t -\frac{1}{2}\bigg(G_\downarrow(X) (\sigma_+\sigma_--\|\sigma_-\psi\|^2) \psi\nonumber\\&+G_\uparrow(X) (\sigma_-\sigma_+-\|\sigma_+\psi\|^2) \psi\bigg)
\end{align} is the continuous part of the stochastic differential equation for the qubit \eqref{eq:quj}.
After partial integration, equation \eqref{eq:bounds} becomes
\begin{align}
&\frac{\diff}{\diff t}\mathsf{E}(f)\nonumber\\&=\int_0^{+\infty}\diff X \int \mathrm{D}\psi \mathrm{D}\psi^*f(\psi,\psi^*,X)\bigg( \mathcal{L}(P)(\psi,\psi^*,X)\nonumber\\&+\partial_\psi(K(\psi) P)(\psi,\psi^*,X)+\partial_{\psi^*} (K^*(\psi)P)(\psi,\psi^*,X)\nonumber\bigg)\\&+\int_0^{+\infty}\diff X \int \mathrm{D}\psi \mathrm{D}\psi^*f(\psi,\psi^*,X)\nonumber\\&\times\int_0^{+\infty}\diff X' \int \mathrm{D}\bar{\psi} \mathrm{D}\bar{\psi}^*\bigg[W(\psi,X|\bar{\psi},X')P(\bar{\psi},\bar{\psi}^*,X',t)\nonumber\\&-W(\bar{\psi},X'|\psi,X)P(\psi,\psi^*,X,t)\bigg]
 +\textrm{Boundary terms.}
\end{align}
With
\begin{align}
&W(\psi,X|\psi',X')\nonumber\\&=G_\uparrow(X')\|\sigma_+\psi'\|\delta\bigg(\frac{\sigma_+\psi'}{\|\sigma_+\psi'\|}-\psi\bigg)\delta(X'-\hbar\omega/\gamma-X)\nonumber\\&+G_\downarrow(X')\times\|\sigma_-\psi'\|\delta\bigg(\frac{\sigma_-\psi'}{\|\sigma_-\psi'\|}-\psi\bigg)\delta(X'+\hbar\omega/\gamma-X)
\end{align}
and 
\begin{align}
&\textrm{Boundary terms}=\frac{\Sigma V}{N\gamma}(T_p^5-X^{5/2})P f\bigg|_0^{+\infty}\nonumber\\&\quad-\frac{10\Sigma VT_p^6}{2N\gamma}f\partial_X P\bigg|_0^{+\infty} +\frac{10\Sigma VT_p^6}{2N\gamma}P\partial_X f\bigg|_0^{+\infty}\nonumber\\&+\int\textrm{D}\psi\textrm{D}\psi^*\int_0^{\frac{\hbar\omega}{N\gamma}}\diff X\,\bigg[f(\phi_+,\phi_+^*,-X)\|\sigma_+\psi\|^2\nonumber\\&\times G_\uparrow\left(\frac{\hbar\omega}{N\gamma}-X\right) P\left(\psi,\psi^*,\frac{\hbar\omega}{N\gamma}-X\right)\nonumber\\&-f(\phi_-,\phi_-^*,X)\|\sigma_-\psi\|^2\nonumber\\&\times G_\downarrow\left(X-\frac{\hbar\omega}{N\gamma}\right)P\left(\psi,\psi^*,X-\frac{\hbar\omega}{N\gamma}\right)\bigg]
\end{align}
At infinity the first three boundary terms drop since we assume that the probability and its derivative are zero at infinity. At $X=0$, the first two terms cancel each other out due to the probability current being zero at the reflective boundary. The third term is zero as well, due to reflective boundary conditions we consider functions which have 0 derivative at $X=0$. The first term in the integral is zero due to the theta function and the second term in the integral is zero since $P(\psi,\psi^*,X<0)=0$.
\section{Hybrid master equation}\label{ap:lin}
Let us define the function
\begin{equation}\label{eq:diffF}
f(\psi,\psi^*,\xi)=\psi\,\psi^* \delta(\xi-X)
\end{equation}
Taking the average $\mathsf{E}(\,.\,)$ of this equation gives the density operator as defined in equation \eqref{eq:rho}
\begin{equation}
\mathsf{E}(f(\psi,\psi^*,\xi))=\rho(X,t).
\end{equation}
The differential of $f$ is
\begin{align}
&\diff f(\psi,\psi^*,\xi)\nonumber\\&= f(\psi+\diff \psi,\psi^*+\diff \psi^*,\xi+\diff \xi)-f(\psi,\psi^*,\xi)  \nonumber\\
&=\sum^{\infty}_{\substack{p=1\\p=k_1+k_2+k_3}}\frac{(\diff \xi)^{k_1}(\diff \psi^*)^{k_2}(\diff \psi)^{k_3}}{k_1!k_2!k_3!}
\partial_{\xi}^{k_1}\partial_{\psi^*}^{k_2}\partial_{\psi}^{k_3}f(\psi,\psi^*,\xi)
\end{align}
To proceed with the calculation, we use the explicit expressions \eqref{eq:quj} and \eqref{eq:tempj} of the differentials. We simplify the above equation by making use of the rules of stochastic calculus, see e.g. \cite{Kurt}. From stochastic calculus it follows that $\diff w^2(t)=\diff t$, $\diff w(t)\diff N_i=0$, and $\diff N_i \diff N_j=\delta_{i,j}\diff N_i$. We thus get the It\^{o}-Poisson stochastic differential 
\begin{align}\label{ap:lin1}
&\diff f(\psi,\psi^*,\xi)\nonumber\\&=(\mathcal{L}^{(1)}_\xi)^\dagger f(\psi,\psi^*,\xi)\diff t+(\mathcal{L}^{(2)}_\xi)^\dagger f(\psi,\psi^*,\xi)\diff t\nonumber\\&+\frac{\sqrt{10\Sigma Vk_B}T_p^3}{\gamma}\partial_\xi f(\psi,\psi^*,\xi)\diff w(t)\nonumber\\
&-\frac{i}{\hbar}\bigg(K(\psi)\partial_\psi-K(\psi)\partial_{\psi^*}\bigg)f(\psi,\psi^*,\xi)\diff t \nonumber\\
&+\bigg(f\left(\phi_+,\phi_+^*,\xi-\frac{\hbar\omega}{N\gamma}\right)-f(\psi,\psi^*,\xi)\bigg)\diff N_\uparrow\nonumber\\&+\bigg(f\left(\phi_-,\phi_-^*,\xi+\frac{\hbar\omega}{N\gamma}\right)-f(\psi,\psi^*,\xi)\bigg)\diff N_\downarrow,
\end{align}
where $\phi_\pm$ are the eigenstates of $\sigma_z$ and $K({\psi})$ \eqref{eq:ok} is the continuous part of the qubit stochastic differential equation \eqref{eq:quj}.
Taking the average $\mathsf{E}(\,.\,)$ of equation \eqref{ap:lin1} we can simplify the equation. The term proportional $\diff w(t)$ cancels due to the It\^{o} description \cite{Kurt}.  Using the definition of $f$ \eqref{eq:diffF} gives the identity \begin{equation}\label{eq:lin2}
\mathsf{E}\big(\psi\psi^*(\mathcal{L}^{(i)}_X)^\dagger\delta(\xi-X)\big)=\mathcal{L}^{(i)}_X\rho(X,t)
\end{equation}
for $i=1,\,2$.
The average of the third line in equation \eqref{ap:lin1} gives
\begin{align}\label{eq:lin3}
&-\frac{i}{\hbar}[H(t),\rho(X,t)]\nonumber\\&+\frac{1}{2}\mathsf{E}\bigg( (G_\downarrow(X) \|\sigma_-\psi\|+ G_\uparrow(X) \|\sigma_+\psi\|)\psi\,\psi^*\bigg|\psi\bigg)\nonumber\\&-\frac{1}{2}\bigg(G_\downarrow(X) \{\sigma_+\sigma_-,\rho(X,t)\}+G_\uparrow(X) \{\sigma_-\sigma_+,\rho(X,t)\}\bigg)
\end{align}
The last two lines become 
\begin{align}\label{eq:lin4}
&G_\downarrow(X) \sigma_-\rho(X,t)\sigma_++G_\uparrow(X) \sigma_+\rho(X,t)\sigma_-\nonumber\\&\quad-\frac{1}{2}\mathsf{E}\bigg( (G_\downarrow(X) \|\sigma_-\psi\|+ G_\uparrow(X) \|\sigma_+\psi\|)\psi\,\psi^*\bigg|\psi\bigg)
\end{align}
Summing equations \eqref{eq:lin2}, \eqref{eq:lin3} and \eqref{eq:lin4} gives \eqref{eq:fullrho}.
\subsection{Discrete hybrid equation}\label{ap:hybrid}
The master equation \eqref{eq:fullrho} is a hybrid master equation. It describes the joint evolution of a classical variable, the temperature of the calorimeter squared $X=T_e^2$, and a quantum variable, the wave function of the qubit. The master equation \eqref{eq:fullrho} can be identified as the hybrid master equation proposed by \cite{ChKoMaSu2011}. 

First let us write the drift-diffusion terms from equation \eqref{eq:fullrho} as discrete jump process
\begin{align}\label{eq:discrho}
&\frac{\diff \rho}{\diff X}(X,t)\nonumber\\&=\sum_{j=\pm}G_j(X-j\Delta X) \rho(X-j\Delta X)-G_j(X) \rho(X)\nonumber\\& -\frac{i}{\hbar} [H(t),\rho(X)]+ G_\downarrow\left(X-\frac{\hbar\omega}{N\gamma}\right)\sigma_-\rho\left(X-\frac{\hbar\omega}{N\gamma}\right)\sigma_+\nonumber \\&+ G_\uparrow\left(X+\frac{\hbar\omega}{N\gamma}\right)\sigma_+\rho\left(X+\frac{\hbar\omega}{N\gamma}\right)\sigma_-\nonumber\\&-\frac{1}{2} G_\downarrow(X)\{\sigma_+\sigma_-,\rho(X)\} -\frac{1}{2} G_\uparrow(X)\{\sigma_-\sigma_+,\rho(X)\},
\end{align}
such that by taking the limit $\Delta X\rightarrow 0$  we retrieve the drift-diffusion process of the temperature squared.
For a set initial temperature, the temperature is thus a discrete variable.

Let us now treat the temperature as a full quantum variable. That is we, expand the Hilbert space of the qubit with an infinite dimensional Hilbert space, which corresponds to the discrete set of temperatures the calorimeter can reach according to equation \eqref{eq:discrho}. Additionally, we define $e_{X,Y}$ as the projector from temperature squared $Y$ to $X$. 
We define the operator $\Phi$ acting on a qubit-temperature squared operator $a$ as
\begin{align}
&\Phi\, a=\sum_{X}\bigg(G_{-}(X) (e_{X+\Delta X,X}\otimes \mathbb{I})a(e^*_{X+\Delta X,X}\otimes \mathbb{I})\nonumber\\&\quad+G_{+}(X) (e_{X-\Delta X,X}\otimes \mathbb{I})a(e^*_{X-\Delta X,X}\otimes \mathbb{I})\nonumber\\&\quad+G_{\uparrow}(X) (e_{X-\hbar\omega/N\gamma,X}\otimes\sigma_-)a(e^*_{X-\hbar\omega/N\gamma,X}\otimes\sigma_+)\nonumber\\&\quad+G_{\downarrow}(X) (e_{X+\hbar\omega/N\gamma,X}\otimes\sigma_+)a(e^*_{X+\hbar\omega/N\gamma,X}\otimes\sigma_-)\bigg).
\end{align}
Following to the results of \cite{ChKoMaSu2011}, this operator is completely positive. Evolution with the adjoint $\Phi^*$ as generator maps states diagonal in $X$ onto states which are diagonal in $X$. A state which is diagonal in $X$ evolves as 
\begin{equation}\label{eq:clasrho}
\rho(t)=\sum_X  e_{X,X}\otimes\rho(X,t).
\end{equation}
where the qubit density at $X$ $\rho(X,t)$ satisfies
\begin{align}
&\frac{\diff \rho(X,t)}{\diff t}=-\frac{i}{\hbar}[H(t),\rho(X,t)]+\sum_Y e_{X,Y}\Phi^*(\rho(t))e^*_{X,Y}\nonumber\\&\quad-\frac{1}{2}\sum_Y\{e_{Y,X}\Phi^*(\mathbb{I}),\rho(X,t)\}
\end{align}
which gives in the limit of $\Delta X\downarrow 0$ gives \eqref{eq:fullrho}.

\section{Effective temperature equation}\label{app:Ef}
We solve equation \eqref{eq:differentorders} at different orders by plugging in the Hilbert expansion \eqref{eq:hilbertexp}. For physical parameters typical for the qubit-calorimeter experiment \cite{PeSo13} the relevant temperature (squared) range is much larger than $\hbar\omega/N\gamma$. For this reason we will treat the rates $G_{\uparrow/\downarrow}(X)$ \eqref{eq:rates2} as differentiable functions and ignore the small jump at $X=\hbar\omega/N\gamma$.
\paragraph{Order $\varepsilon^0$}
The lowest order equation is
\begin{equation}\label{eq:zerotho}
M^{(0)}\vec{P}^{0}(X)=0.
\end{equation}
The zeroth order of the Hilbert expansion \eqref{eq:hilbertexp} can be written into the form 
\begin{equation} 
\vec{P}(X,\tau)=F^{(0)}(X,\tau)\vec{Q}(X,\tau)
\end{equation}
where, taking $G(X)=G_\uparrow(X)+G_\downarrow(X)$,
\begin{align}
&\vec{Q}(X)=\nonumber\\&\frac{1}{G(X)^2 + 8\lambda^2}\begin{pmatrix}
G_\uparrow(X)G(X)+4\lambda^2\\
G_\downarrow(X)G(X)+4\lambda^2\\
-2i\lambda (G_\downarrow(X)-G_\uparrow(X))\\
2i\lambda (G_\downarrow(X)-G_\uparrow(X))
\end{pmatrix}
\end{align}
solves \eqref{eq:zerotho} and satisfies $Q_1(X)+Q_2(X)=1$.

\paragraph{Order $\varepsilon^1$}
The first order correction to $\vec{P}(X,t)$ solves
\begin{align}\label{eq:firsto}
(M^{(0)}\vec{P}^{(1)})(X,\tau)=& \frac{\diff \vec{P}^{(0)}}{\diff \tau}(X,\tau)-(\mathcal{L}^{(1)}_X\vec{P}^{(0)})(X,\tau)\nonumber\\&-\partial_X\big(M^{(1)}\vec{P}^{(0)}\big)(X,\tau).
\end{align}
By Fredholm's alternative \cite{Pavbook}, the above equation is solvable if the solvability condition satisfied. The solvability condition requires that non-homogeneous part of the above equation, i.e. the right hand side, is zero on the kernel of the adjoint of $M^{(0)}$. Concretely, given that the kernel of $(M^{(0)})^\dagger$ is \begin{equation}
v_1=(1,\,1,\,0,\,0),
\end{equation}
the solvability condition requires that
\begin{equation}\label{eq:ff0}
\partial_t F^{(0)}=\mathcal{L}^{(1)}_X F^{(0)} +\langle v_1| \partial_XM^{(1)} (\vec{Q} F^{(0)})\rangle
\end{equation}
should be satisfied. 

The matrix $M^{(0)}$ has eigenvalues 0, $\lambda_1$, $\lambda_2$,  $\lambda_3$, with corresponding right eigenvectors $Q$, $w_2$, $w_3$, $w_4$ and left eigenvectors $v_1$, $v_2$, $v_3$ and $v_4$.
The vector $v_2=(0,0,1,1)$, it is straightforward to see that  $\langle v_2,Q\rangle=0$ and $v_2\in\, \textrm{ker}(M_1^\dagger)$,. Projecting $v_2$ on both sides of equation \eqref{eq:firsto} gives
\begin{equation}\label{eq:v2}
\langle v_2 , M^{(0)} P^{(1)}\rangle=\lambda_2 \langle v_2,P^{(1)}\rangle=0
\end{equation}
For $j=3,\, 4$ we have
\begin{align}\label{eq:v34}
&\lambda_j\langle v_j,P^{(1)}\rangle\nonumber\\&=-\langle v_j| (\partial_X\vec{Q}) f(X) F^{(0)}\rangle +\langle v_j|\vec{Q}\rangle\langle v_1|\partial_X(M^{(1)}F^{(0)})\rangle \nonumber\\&\quad-\langle v_j|\partial_X(M^{(1)}\vec{Q} F^{(0)})\rangle\nonumber\\
&= -\langle v_j| (\partial_X\vec{Q}) f(X) F^{(0)}\rangle  -\langle v_j|\partial_X(M^{(1)}\vec{Q} F^{(0)})\rangle.
\end{align} 
Going to the last line, we used that $\langle v_j,\vec{Q}\rangle=0$ for $j\neq 1$ and $f(X)=-\frac{\Sigma V}{N\gamma}(T_p^5-X^{5/2})$.
Furthermore, the eigenvectors satisfy the completeness relation
\begin{equation}
\mathbb{I}=|Q\rangle\langle v_1|+\frac{|w_2\rangle\langle v_2|}{\langle v_2|w_2\rangle}+\frac{|w_3\rangle\langle v_3|}{\langle v_3|w_3\rangle}+\frac{|w_4\rangle\langle v_4|}{\langle v_4|w_4\rangle}
\label{eq:com2}
\end{equation}
\paragraph{Order $\varepsilon^2$} We get the equation 
\begin{align}\label{eq:secondo}
&(M^{(0)}\vec{P}^{(2)})(X,\tau)= \frac{\diff \vec{P}^{(1)}}{\diff \tau}(X,\tau)-(\mathcal{L}^{(1)}_X\vec{P}^{(1)})(X,\tau)\nonumber\\&-(\mathcal{L}^{(2)}_X\vec{P}^{(0)})(X,\tau)-(M^{(0)}\vec{P}^{(1)})(X,\tau)\nonumber\\&-\partial_X\big(M^{(1)}\vec{P}^{(1)}\big)(X,\tau)-\frac{1}{2}\partial_X\big(M^{(2)}\vec{P}^{(0)}\big)(X,\tau).
\end{align}
By projecting the kernel of $(M^{(0)})^\dagger$ on \eqref{eq:secondo}, we find the second order solvability condition
\begin{align}
\partial_t F^{(1)}=&\mathcal{L}^{(1)}F^{(1)}+\mathcal{L}^{(2)}F^{(2)}+\langle v_1|\partial_XM^{(1)} P^{(1)}\rangle\nonumber\\&+\langle v_1|\partial_X^2M^{(2)} P^{(0)}\rangle/2.
\end{align}
Using the completeness relation \eqref{eq:com2} in the third term on the right hand side, we find
\begin{align}\label{eq:ff1}
&\partial_t F^{(1)}=\mathcal{L}^{(1)}F_1+\mathcal{L}^{(2)}F_0\nonumber\\&+\langle v_1|\partial_X( M^{(1)} Q F^{(1)})\rangle+\langle v_1|\partial_X\left(\frac{|(M^{(1)}w_3\rangle\langle v_3|P^{(1)}\rangle}{\langle v_3|w_\rangle}\right)\nonumber\\&+\langle v_1|\partial_X\left(\frac{|(M^{(1)}w_4\rangle\langle v_4|P^{(1)}\rangle}{\langle v_4|w_4\rangle}\right)+\langle v_1|\partial_X^2(M^{(2)} P^{(0)})\rangle/2
\end{align}
By summing equations \eqref{eq:ff0} and \eqref{eq:ff1}, and using \eqref{eq:v2} and \eqref{eq:v34}, we find the Fokker-Planck equation \eqref{eq:pertF} for $F=F^{(0)}+\varepsilon F^{(1)}$.
\bibliography{lit}
\end{document}